\documentclass[sigconf]{acmart}
\pdfoutput=1

\usepackage{bm}
\usepackage{multirow}
\usepackage{subfigure}
\usepackage{algorithmic}
\usepackage[ruled, linesnumbered]{algorithm2e}
\usepackage{booktabs} 
\usepackage{amsmath}

\usepackage{amssymb}
\usepackage{amsmath}
\usepackage{xcolor}
\usepackage{multirow}
\usepackage{graphicx}
\usepackage{subfigure}
\usepackage{amsfonts}
\usepackage{balance}

\AtBeginDocument{%
  \providecommand\BibTeX{{%
    \normalfont B\kern-0.5em{\scshape i\kern-0.25em b}\kern-0.8em\TeX}}}

\newcommand{\model}{\emph{PGD}}
\newcommand{\fmodel}{\emph{privileged graph distillation model~(PGD)}}

\copyrightyear{2021}
\acmYear{2021}
\setcopyright{acmlicensed}
\acmConference[SIGIR '21]{Proceedings of the 44th International ACM SIGIR Conference on Research and Development in Information Retrieval}{July 11--15, 2021}{Virtual Event, Canada}
\acmBooktitle{Proceedings of the 44th International ACM SIGIR Conference on Research and Development in Information Retrieval (SIGIR '21), July 11--15, 2021, Virtual Event, Canada}
\acmPrice{15.00}
\acmDOI{10.1145/3404835.3462929}
\acmISBN{978-1-4503-8037-9/21/07}

\begin{document}
\fancyhead{}

\title{Privileged Graph Distillation for Cold Start Recommendation 
}

\author{Shuai Wang$^{1,2}$, Kun Zhang$^{1,2,*}$, Le Wu$^{1,2,3}$, Haiping Ma$^{4}$, Richang Hong$^{1,2}$, Meng Wang$^{1,2,3}$}
\affiliation[obeypunctuation=true]{\institution{$^1$ Key Laboratory of Knowledge Engineering with Big Data, Hefei University of Technology, \country{China}}}
\affiliation[obeypunctuation=true]{\institution{$^2$ School of Computer Science and Information Engineering, Hefei University of Technology, \country{China}}}
\affiliation[obeypunctuation=true]{\institution{$^3$ Institute of Artificial Intelligence, Hefei Comprehensive National Science Center, \country{China}}}
\affiliation[obeypunctuation=true]{\institution{$^4$ Anhui University, \country{China}}}
\email{{wangshuai418520, zhang1028kun, lewu.ustc, hongrc.hfut, eric.mengwang}@gmail.com, hpma@ahu.edu.cn}
\thanks{Kun Zhang is the corresponding author.}
 
\begin{abstract}
The cold start problem in recommender systems is a long-standing challenge, which requires recommending to new users (items) based on attributes without any historical interaction records. In these recommendation systems, warm users~(items) have privileged collaborative signals of interaction records compared to cold start users~(items), and these Collaborative Filtering (CF) signals are shown to have competing performance for recommendation. Many researchers proposed to learn the correlation between collaborative signal embedding space and the attribute embedding space to improve the cold start recommendation, in which user and item categorical attributes are available in many online platforms. 
However, the cold start recommendation is still limited by two embedding spaces modeling and simple assumptions of space transformation. As user-item interaction behaviors and user~(item) attributes naturally form a heterogeneous graph structure, in this paper, we propose a ~\fmodel. The teacher model is composed of a heterogeneous graph structure for warm users and items with privileged CF links. The student model is composed of an entity-attribute graph without CF links. Specifically, the teacher model can learn better embeddings of each entity by injecting complex higher-order relationships from the constructed heterogeneous graph. The student model can learn the distilled output with privileged CF embeddings from the teacher embeddings. 
Our proposed model is generally applicable to different cold start scenarios with new user, new item, or new user-new item.
Finally, extensive experimental results on the real-world datasets clearly show the effectiveness of our proposed model on different types of cold start problems, with average $6.6\%, 5.6\%, $ and $17.1\%$ improvement over state-of-the-art baselines on three datasets, respectively.

\end{abstract}



\maketitle

\section{Introduction}
Collaborative Filtering (CF) is widely applied to various scenarios of recommender systems, which provides personalized item recommendation based on past user behaviors, such as purchasing a product~\cite{koren2009matrix,salakhutdinov2008bayesian,rendlebayesian}.
Recently, graph based recommendations have shown huge success for solving data sparsity problem~\cite{berg2017graph,wang2019neural,wu2019neural}.
Since the user-item interactions naturally form a graph, graph based recommendations obtain better user and item representations by aggregating higher-order neighbor information in a data sparsity setting.
However, the cold start problem is still a challenge in CF based recommendation.
Since new users or items have no historical interaction records, a 
conventional way to solve the cold start problem is to introduce additional data such as reviews, social networks, attributes, etc.
Among them, user and item attributes are easily acquired in most online platforms (e.g., Facebook, Amazon) and described specific features.
In this paper, we focus on attribute information in the cold start setting.

For most attribute enhanced recommendation methods, we summarize them into three
categories according to the difference of input data: CF-based, content-based, and hybrid methods. 
Given the history interaction data and attributes, some researchers leverage collaborative information of the existing entities and the attribute similarity for new user (item) recommendations ~\cite{goldberg2001eigentaste,zhou2011functional,sedhain2014social}.
However, they do not model attribute information to feature space.
Deep neural networks have achieved better performance in feature engineering modeling.
Content-based methods make full use of auxiliary information of users and items to enhance the modeling of preference embedding~\cite{gantner2010learning,van2013deep,lian2018xdeepfm,wang2015collaborative,cheng2021long}. For example, DeepMusic~\cite{van2013deep} and CDL~\cite{wang2015collaborative} were proposed to incorporate content data into deep neural networks and learned a general transformation function for content representations. 
A simple assumption is that the attribute information can be mapped into the embedding space by a general transformation function, which ignores collaborative signals for new users or items side.
In order to overcome this shortcoming and further improve the model performance based on the content information,
hybrid methods are proposed. 
Hybrid models fuse the CF and content embedding, and model the relations between CF and content space~\cite{volkovs2017dropoutnet,zhu2020recommendation}.
For example, DropoutNet~\cite{volkovs2017dropoutnet} was proposed to make full use of content and pretrained CF embedding for recommendation.
However, most of these methods still have some weaknesses in dealing with those new users (items), that have no interactions with existing items (users).

Graph based recommendations are limited by user-item links.
To obtain unseen node embedding in a graph, inductive representation learning combines node features and graph structures for node embedding ~\cite{hamilton2017inductive,ying2018graph,wu2020learning}.
For example, PinSage is a content-based Graph Convolutional Networks (GCN) model for recommending items, which gathers both graph structure and node features for embedding learning~\cite{ying2018graph}.
They still have weaknesses in tackling new user (item) problem mentioned above. 
In other words, how to make recommendations for new users (items), who have no links during test, is still challenging.  
Since user-item links are available during train while not available during test,
interaction data is capable of providing privileged information.
This problem can also be treated as how to leverage attribute information to distill privileged information for better recommendations of new users (items).

To this end, in this paper, we take advantages of graph learning and knowledge distillation in privileged information modeling and propose a novel \fmodel~for the cold start problem, which new users (items) have no link during test. 
Specifically, we introduce attributes of users (items) as nodes into a user-item graph and construct a heterogeneous graph, so that attribute representations can capture higher order information
during embedding propagation.
Since privileged information is only available offline and effective for prediction, we employ knowledge distillation method to tackle the cold start problem. 
More specifically, the teacher model can access all the information and make full use of attributes for privileged information learning and user preference modeling. 
The student model is constructed on an entity-attribute graph without CF links, which can obtain privileged information based on attributes under the guidance of the teacher model. 
Then, the student model can fuse CF signals of user or item embedding for final recommendations. 
Thus, \model~can not only make full use of attribute information for a better recommendation, but also alleviate the cold start problem when recommending for new users or items. 
Finally, we detail the cold start problem in recommendation into three sub-tasks and evaluate the model performance with three datasets. 
Extensively experimental results demonstrate the superiority of our proposed \model.

\section{Related Work}

\subsection{Cold Start Recommendation}
CF-based algorithms personally recommend products by collecting explicit rating records and implicit feedback, which are widely applied in various recommendation
systems~\cite{koren2009matrix,salakhutdinov2008bayesian,rendlebayesian}.
These methods leverage matrix factorization to obtain low-dimensional representations of users and items. 
For example, Salakhutdinov et al.~\cite{rendlebayesian} proposed Bayesian Personalized Ranking (BPR), which learned user and item latent vectors based on implicit feedback.
Moreover, with the development of GCN, plenty of GCN-based CF methods are proposed to learn better collaborative filtering and alleviate the data sparsity problem~\cite{chen2020revisiting,wu2019neural,he2020lightgcn}. 
For example, Chen et al.~\cite{chen2020revisiting} designed LR-GCCF model to simplify the embedding propagation process with linear graph convolutions, which achieved excellent performance. 
However, most CF-based methods require links between users and items, which limit their applications. 
In order to solve the cold start problem, CF-based methods leverage social data and basic matrix factorization to capture the new users' preferences conventionally~\cite{goldberg2001eigentaste,zhou2011functional,ren2017social,sedhain2014social}.
Social data based methods first keep the pretrained CF representations on implicit feedback data, and then generate the new user's embedding with the connection between new users and old users~\cite{sedhain2014social}.
Despite the achievements they have made, most of these models still have some drawbacks.
These methods cannot be widely used in the case of both new users and new items, and underestimate the potential of users' and items' attribute information .

In order to remedy the shortcomings of CF-based methods, researchers proposed to utilize additional content information and designed content-based methods. 
Content-based methods take the profile as input and train a general transform function for content information, in which new user or item representation can be generated.
These methods usually learn a mapping function to transform the content representation into collaborative space~\cite{gantner2010learning,wang2015collaborative,van2013deep}, and leverage deep cross-network structure to capture higher-order relationships between features~\cite{wang2017deep,lian2018xdeepfm}. 
For example, xDeepFM was proposed to model cross interactions at the vector-wise level explicitly~\cite{lian2018xdeepfm}. 
In order to solve the cold start problem in graph based recommendations, PinSage was proposed to leverage both attributes as well as the user-item graph structure to generate better embeddings~\cite{ying2018graph}.
However, most of these methods do not consider the complicated connection between CF embedding space and content space for each user (item), in which new user (item) 
representations cannot reflect the association with CF information.

To make full use of both CF-based methods and content-based methods, hybrid models are proposed to make better recommendations~\cite{volkovs2017dropoutnet,zhu2020recommendation,wu2020learning}. 
Most of these methods learn CF embedding and transformation functions to minimize prediction errors.
A typical example is Heater ~\cite{zhu2020recommendation}, which dropped CF signals randomly to imitate new users or items situations.
In particular, the CF representation is pretrained as a constraint for content embedding learning.
The final prediction is conducted with a random choice of CF representation or content representation.
Since the construction of user-item bipartite graph relies on interaction records, the learning of new user (item) representation is still a problem in graph based recommendations.
Thus, inductive learning methods of graph are proposed to tackle unseen nodes' representation problem~\cite{hamilton2017inductive,chami2019hyperbolic,zeng2019graphsaint,zhang2019inductive}. 
Among these methods, TransGRec was proposed to feed the item's CF information and content information into the node initialization layer of the graph~\cite{wu2020learning}.
Especially, TransGRec was designed to learn graph's structure information with the transfer network which is used to solve new user (item) problem.

\subsection{Knowledge Distillation and Applications in Recommendations}
Knowledge distillation is first proposed to address the lack of data and devices with limited resources.
It aims to learn a better student model from a large teacher model and abandon the teacher model at the testing stage. 
In recent years, the knowledge distillation is presented in three ways: logits output~\cite{hinton2015distilling,mirzadeh2020improved,zhou2018rocket}, intermediate layers~\cite{romero2014fitnets,zagoruyko2016paying}, and relation-based distillation~\cite{park2019relational,chen2020learning,peng2019correlation,liu2019knowledge}.
Most of methods assume that the teacher model and the student model input the same regular data in the distillation process, which means the available information at test is the same as at train.
In the real world, some information is helpful for prediction tasks but not always available, 
which called privileged information (e.g., medical reports in pathology analysis).
Therefore, privileged distillation is proposed to tackle the lack of data problem in testing online, in which privileged information is only fed into the teacher model.
Lopez et al.~\cite{lopez2015unifying} proposed an approach that guided the student model with fewer data and distilled the teacher model's privileged information. 
Since knowledge distillation is capable to solve the data missing and time-consuming problems,
it attracts attention in recommendation areas.
There are some works that get light models with better performance by model distillation~\cite{tang2018ranking,zhang2020distilling,wang2020next,kang2020rrd}, which solve the problem of limited equipment resources and reduce the running time.
For example, Zhang et al.~\cite{zhang2020distilling} constructed an embedding based model to distill user's meta-path structure and improve accuracy and interpretability.
Meanwhile, to solve the problem that privileged information is unavailable in online recommendations, researchers proposed to introduce privileged distillation into recommendations~\cite{chen2018adversarial,xu2019privileged}.
Selective Distillation Network ~\cite{chen2018adversarial} was proposed to use a review process framework as the teacher model, so that the student model can distill effective review information.
Xu et al.~\cite{xu2019privileged} proposed Privileged Features Distillation (PFD) to distill privileged features
and in click-through rate and achieved better performance in click-through rate and conversion rate.
However, most methods haven't addressed the new user or item problem.

In this paper, we treat interaction data as privileged information and design student network to imitate the situation of new users or items. Our goal is to improve model performance on cold start problems by distilling teacher's graph structure information and privileged information.

\section{Problem Definition}
\label{s:problem}

In a collaborative filtering based recommendation system, there are two sets
of entities: a userset~{\small$U$~($|U|\!=\!M$)}, and an itemset~{\small$V$~($|V|\!=\!N$)}. Since implicit feedback is available
in most scenarios, we use a rating matrix {\small$\mathbf{R}\in\mathbb{R}^{M\times N}$} to denote the interaction information,
with $r_{ij} = 1$ indicates observed interaction between user $i$ and item $j$,
otherwise it equals to 0. 
Traditionally, the user-item interaction behavior could be naturally formulated
as a user-item bipartite graph: $\mathcal{G}_R=<U \cup V, \mathbf{A}^R>$, where
the graph adjacent matrix is constructed from the interaction matrix {\small$\mathbf{R}$}:

\begin{small}
\begin{flalign}\label{eq:adj_matrix1}
\mathbf{A}^R=\left[\begin{array}{cc}
\mathbf{0}^{M\times M} & \mathbf{R}\\
\mathbf{R}^T & \mathbf{0}^{N\times N}
\end{array}\right].
\end{flalign}
\end{small}

Most of the attributes are sparse and categorical, and we generally convert continuous attributes to discrete distributions.
Meanwhile, the entity attribute matrix {\small$\mathbf{X}\in\mathbb{R}^{(M+N)\times D}$} is usually treated as the supplement information for user-item bipartite graph, where {$D$} is the dimension of user and item attributes. Besides, we employ $\mathbf{x}_i\in\mathbb{R}^D$ and $\mathbf{x}_{M+j}\in\mathbb{R}^D$
to denote the $i^{th}$ user one-hot attribute and the $j^{th}$ item one-hot attribute
~{($0 \leq i \textless \small{M}$~, $0 \leq j \textless  \small{N}$)}.
For $\mathbf{x}_i$, the attribute's indices are between $0$ and $(D_u-1)$.
For $\mathbf{x}_j$, the attribute's indices are between $D_u$ and $(D-1)$,
where $D_u$ is the dimension of user attributes.

The goal of graph based recommendations is to measure the user preference and predict the preference score matrix {\small$\hat{\mathbf{R}}\in\mathbb{R}^ {M\times N}$}. 
In order to evaluate the model performance, we also split the recommendation task into three sub-tasks to analyze the real-world scenarios in a detailed way. 
\begin{itemize}
    \item[\textit{Task 1}:]  When a new user with attributes appears, we recommend existing~(old) products to new users;
    
    \item[\textit{Task 2}:]  When a new product with attributes appears, 
    we have to recommend new products to existing~(old) users; 
    
    \item[\textit{Task 3}:] When new users and new products appear at the same time, we have to recommend new products to new users.
\end{itemize}

To this end, we propose a novel \fmodel\ to tackle the above challenges.
Next, we will introduce the technical details of \model.

\section{The Proposed Model}
\label{s:model} 
Figure~\ref{fig:framework} illustrates the overall architecture of our proposed \model, which consists of three main components: 
1) \textit{Teacher model}: leveraging existing user-item interactions to learn user preference representation and item representation;
2) \textit{User Student model}: focusing on new user preference modeling;
3) \textit{Item Student model}: concentrating on new item modeling.

Before introducing the technical details, we first introduce the necessary notations for the sake of convenience. 
We use {\small$\mathbf{U}\in\mathbb{R}^{M\times d}$} and {\small$\mathbf{V}\in\mathbb{R}^{N\times d}$} to denote the free embedding matrix of user and item respectively, where  $M$ and $N$ represent the number of users and items. $d$ is the dimension of free embedding. 
Moreover, we leverage {\small$\mathbf{Y}\in\mathbb{R}^{D\times d}$} to represent the user attribute and item attribute node embedding matrix.
Besides, we employ $\mathbf{y}_k$ and $\mathbf{y}_l$ to denote the $k^{th}$ user attribute and the $l^{th}$ item attribute~{($0 \leq k \textless D_u$~, $D_u \leq l \textless D$)}. 
Next, we will introduce the technical details of our proposed \model.

\begin{small}
\begin{figure*} [htb]
  \begin{center}
    \includegraphics[width=0.9\textwidth]{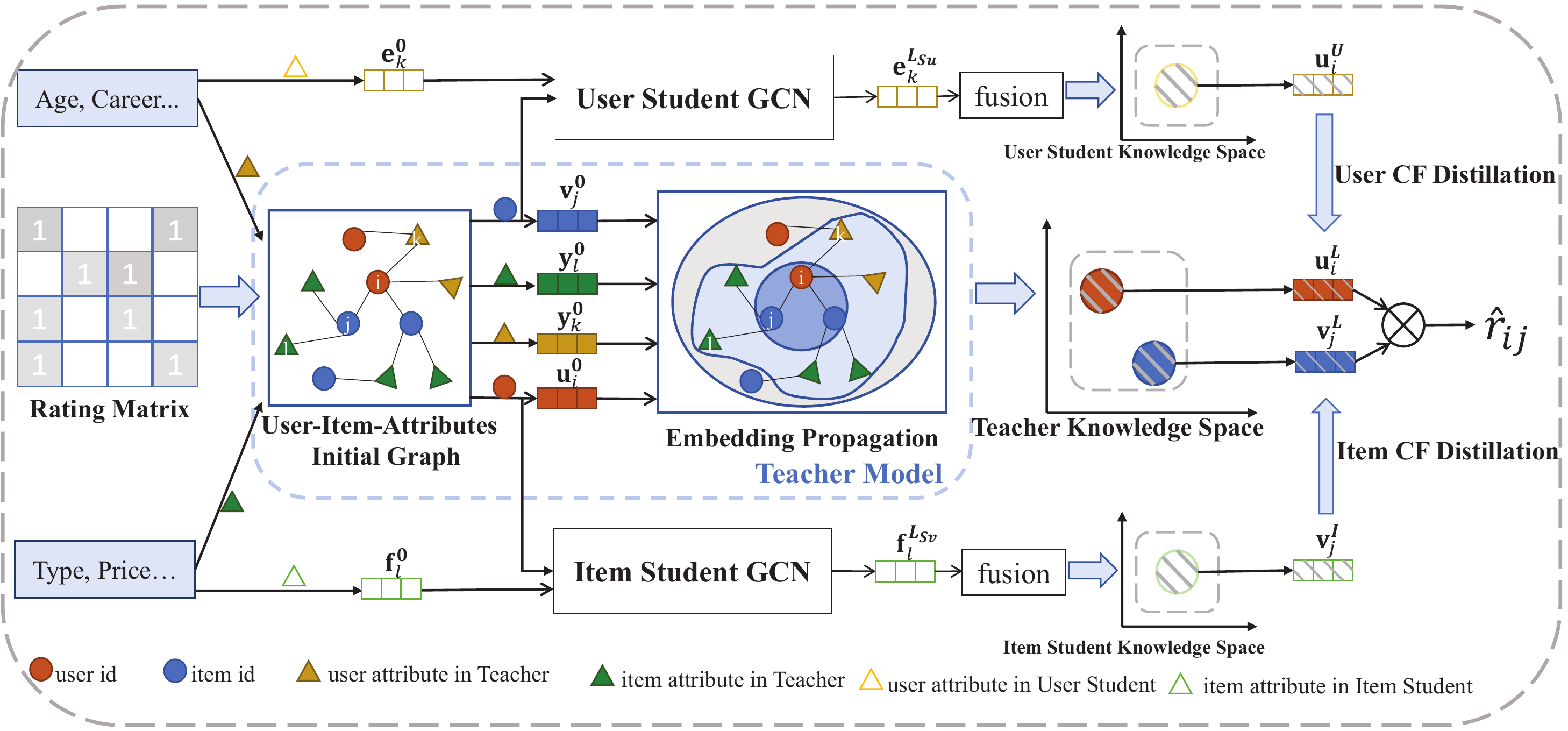}
  \end{center}
    \vspace{-0.3cm}
  \caption{The overall framework of our proposed model.}\label{fig:framework}
  \vspace{-0.3cm}
\end{figure*}
\end{small}

\begin{small}
\begin{figure} [htb]
  \begin{center}
    \includegraphics[width=0.4\textwidth, height=60mm]{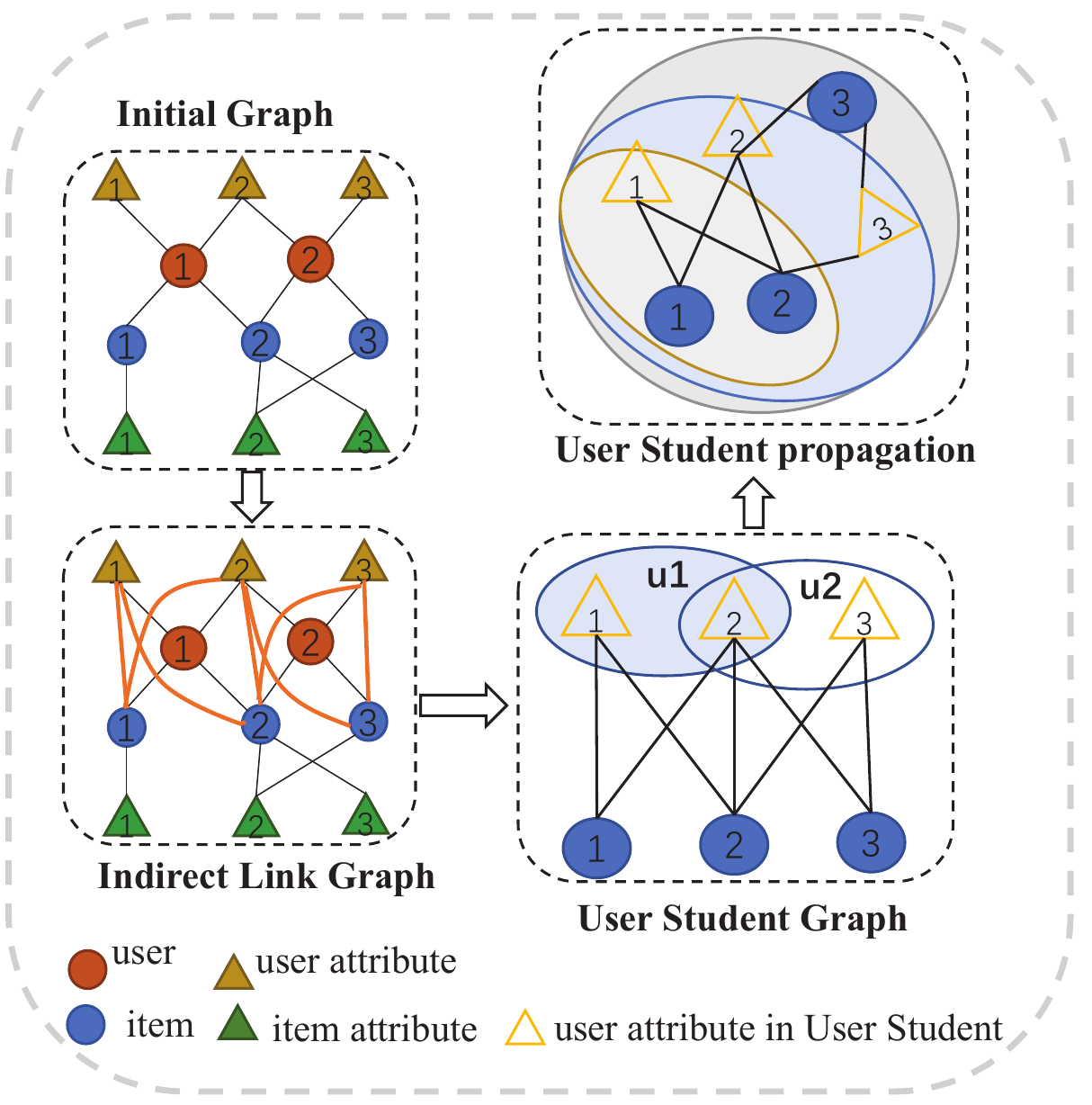}
  \end{center}
    \vspace{-0.5cm}
  \caption{The user student framework of \model.}
  \vspace{-0.3cm}
  \label{fig:student}
\end{figure}
\end{small}

\subsection{Teacher Model}
\label{s:teacher_model}
As mentioned before, we intend to leverage attribute information to build connections for new users and new items. 
To this end, we construct a novel graph with the attributes as the nodes, and design a novel GCN, which we name as \textit{Teacher model}, to generate comprehensive user and item embeddings, as well as predict the ratings of users to items. The teacher model's structure could be formulated
as a user-item-attributes graph:
$\mathcal{G}=<U \cup V \cup \mathbf{X}, \mathbf{A}>$, where the graph matrix
is constructed from the rating adjacent matrix $\small\mathbf{A}^R$ and attribute matrix $\small\mathbf{X}$:

\begin{small}
\begin{flalign}\label{eq:adj_matrix2}
\mathbf{A}=\left[\begin{array}{cc}
\mathbf{A}^R & \mathbf{X}\\
\mathbf{X}^T & \mathbf{0}^{D\times D}
\end{array}\right],
\end{flalign}
\end{small}

Next, we first introduce the graph construction and model initialization. Then, we give a detailed description of the embedding propagation and model prediction.

\textbf{Model Initialization Layer. } 
In this layer, we leverage the free embedding matrix {\small$\mathbf{U}\in\mathbb{R}^{M\times d}$} and {\small$\mathbf{V}\in\mathbb{R}^{N\times d}$} to denote users and items.
The attribute embeddings of users and items are represented with {\small$\mathbf{Y}$}.
They are treated as input and initialized with Gaussian Distribution, then updated during the propagation of GCN. 
We have to note that the free embedding matrix {\small$\mathbf{U}$}, {\small$\mathbf{V}$} will be shared with \textit{Student model}, which will be introduced in the following parts. 
    
\textbf{Embedding Propagation Layer.} 
In this part, we employ GCN to propagate users' (items', user attributes', item attributes') embeddings to capture higher-order information and obtain the proximity between four different type nodes for better node representation. 
Specifically, let $\mathbf{u}_i^t$ and $\mathbf{v}_j^t$ denote user $i$'s embedding and item $j$'s embedding at $t^{th}$ layer. 
And, $\mathbf{y}_k^{t}$ denotes the attribute embedding for user, 
$\mathbf{y}_l^{t}$ denotes the attribute embedding for item.
We leverage the output of Initial Embedding Layer as the initial input of this layer, which means $\mathbf{u}_i^0 = \mathbf{u}_i$, $\mathbf{v}_j^0 = \mathbf{v}_j$, $\mathbf{y}_{k}^{0} = \mathbf{y}_{k}$, $\mathbf{y}_{l}^{0} = \mathbf{y}_{l}$.

In order to extract the node embedding at $(t+1)^{th}$ with the consideration of its neighbors' embeddings and its own free embedding at the $t^{th}$ layer, we utilize the graph propagation and pooling operation to update the embedding of each node. 
Taking user $i$ as an example, we leverage $A_i = \{j|r_{ij} = 1\}\cup \{k|x_{ik} = 1\}$ to denote the item set that he has clicked and his corresponding attribute set. 
The updating process can be formulated as follows:
\begin{flalign}
\label{eq:embedding propagation1}
\mathbf{u}_i^{t+1} &= (\mathbf{u}_i^{t}+
\sum_{j\in {A}_i} \frac{\mathbf{v}_j^{t}}{|{A}_i|}+
\sum_{k\in {A}_i} \frac{\mathbf{y}_k^{t}}{|{A}_i|}).
\end{flalign}
By employing this layer, 
\model~not only utilizes item neighbor information to describe the user's implicit
preference, but also makes full use of attributes for the user's explicit feature.

Similarly, \model~is capable of updating the item embedding based on users who have clicked it and the corresponding item attributes.
Therefore, we leverage 
$A_{M+j} = \{i|r_{ij} = 1\} \cup \{l|x_{(j+M)l} = 1\}$ to denote the user set who has clicked the item $j$ and the corresponding attribute set of item $j$. 
Then, the updating operation for item $j$ in the $(t+1)^{th}$ layer can be described as follows:
\begin{flalign}\label{eq:embedding propagation2}
\mathbf{v}_j^{t+1} &= (\mathbf{v}_j^{t}+
\sum_{i\in {A}_{M+j}} \frac{\mathbf{u}_i^{t}}{|{A}_{M+j}|}+
\sum_{l\in {A}_{M+j}} \frac{\mathbf{y}_l^{t}}{|{A}_{M+j}|}).
\end{flalign}
Besides, we add attribute nodes in GCN to enhance user preference modeling. 
Thus, user attribute embedding can be updated based on all users who have the same attributes.
Meanwhile, the item attribute embedding can be updated in a similar way. 
The updating process at the $(t+1)^{th}$ layer can be formulated as follows:
   \begin{align}
   \begin{split}
    \mathbf{y}_k^{t+1} &= \mathbf{y}_k^{t}+
    \sum_{i\in {A}_{k+M+N}} \frac{\mathbf{u}_i^{t}}{|{A}_{k+M+N}|}, 0 \leq  k \textless D_u, \\
    \mathbf{y}_l^{t+1} &= \mathbf{y}_l^{t} + \sum_{j\in {A}_{l+M+N}} \frac{\mathbf{v}_j^{t}}{|{A}_{l+M+N}|},  D_u \leq l \textless D, 
   \end{split}
   \end{align}
where $A_{k+M+N}=\{i|x_{ik}=1\} \in {\small\mathbf{X}}$ denotes the user set who has the attribute $y_k$. 
$A_{l+M+N}=\{j|x_{(j+M)l} = 1\} \in {\small\mathbf{X}}$ denotes the item set that has the attribute $y_l$.

In order to illustrate the embedding propagation process more clearly, we formulate the fusion embedding in the matrix norm. Let matrix {\small$\mathbf{U}^t$}, {\small$\mathbf{V}^t$}, {\small$\mathbf{Y}^t$} denote the embedding matrices of users ,items and attributes after $t^{th}$ propagation, then the updated embedding matrices after $(t+1)^{th}$ propagation as:

\begin{small}
\begin{flalign}\label{eq:new GCN}
\left[\begin{array}{c}
\mathbf{U}^{t+1} \\
\mathbf{V}^{t+1} \\
\mathbf{Y}^{t+1} \end{array}\right]
=(\left[\begin{array}{c}
\mathbf{U}^{t} \\
\mathbf{V}^{t} \\
\mathbf{Y}^{t}\end{array}\right]+
\mathbf{D}^{-1}\mathbf{A}\times
\left[\begin{array}{c}
\mathbf{U}^{t} \\
\mathbf{V}^{t}\\
\mathbf{Y}^{t}\end{array}\right]),
\end{flalign}
\end{small}
where {\small$\mathbf{D}$} is the degree matrix of {\small$\mathbf{A}$}, which could efficiently propagate neighbors' embeddings and update fusion matrices.

\textbf{Model Prediction Layer.}
In this layer, we treat the output of Embedding Propagation Layer as the final user embedding $\hat{\mathbf{u}}_i$ and item embedding $\hat{\mathbf{v}}_j$. 
In this layer, we treat the output of Embedding Propagation Layer as the final user embedding and item embeddings (i.e., $\mathbf{u}_i^L, \mathbf{v}_j^L$), where $L$ is the number of GCN layers in Teacher model. Then, we predict user i's rating to item j by calculating the dot product of their embeddings, which can be formulated as follows:
\begin{align}
\label{eq:teacher_predict}
\hat{r}_{ij} = \hat{\mathbf{u}}_i(\hat{\mathbf{v}}_j)^T =  \mathbf{u}_i^L(\mathbf{v}_j^L)^T.
\end{align}

\subsection{Student Model}
As mentioned before, we introduce attribute information of users and items to alleviate the cold start problem in GCN based recommendation. 
However, attribute information still has some weaknesses in analyzing the collaborative filtering information of users, which is very important for user preference modeling. 
To this end, we intend to leverage distillation techniques to train a student model, which can utilize the attribute information to access the collaborative signals in the teacher model. 
Along this line, the student model can make full use of attribute information to model user preference comprehensively. 
In concerned details, the student model can be classified into two sub-models based on attribute source~(i.e., user attributes or item attributes): 1) \textit{User Student model}, 2) \textit{Item Student model}.
Specially, the attribute embedding of users in \textit{User Student model} represented with {\small$\mathbf{E}=\{\mathbf{e}_0, \mathbf{e}_1,..., \mathbf{e}_{(D_u-1)}\}$} and the attribute embedding of items in \textit{Item Student model} represented with {\small$\mathbf{F}=\{\mathbf{f}_{D_u}, \mathbf{f}_{D_u+1},..., \mathbf{f}_{(D-1)}\}$}.
The former focuses on the new user problem and takes user attributes and items as input. 
The latter focuses on the new item problem and takes item attributes and users as input. 
The framework is illustrated in Figure~\ref{fig:student}.
Since these two sub-models perform in a similar way, we take the \textit{User Student model} as an example to introduce the technical details for the sake of simplicity in the following parts. 

\textbf{Graph Construction.} 
Since the direct connections between new users and items are unavailable in the student model, we first need to construct the graph between new users and items based on the attribute information. 
As illustrated in Figure~\ref{fig:student}, if user $i$ has clicked the item $j$, we could obtain the direct link between user $i$ and item $j$ in the teacher graph. However, this direct link is unavailable in the student graph. 
To this end, we employ indirect links between user attributes and items to replace the direct link between user and items. 
Specifically, if user $i$ have clicked item $j$, which will not be provided to the student model, we link the attributes of user $i$ to item $j$ to construct the user-\textit{attribute}-item graph for \textit{User Student model}. 
Moreover, if multiple users with attribute $k$ have clicked item $j$, we will assign a higher weight to the indirect link between attribute $k$ and item $j$.

We employ {\small$\mathbf{S}_u\in\mathbb{R}^{N\times D_u}$} to denote item-user attribute matrix and {\small$\mathbf{S}_v\in\mathbb{R}^{M\times D_v}$}
denote user-item attribute matrix, where {\small$\mathbf{S}_u$} and {\small$\mathbf{S}_v$}
is constructed from the user-item graph adjacent matrix $\mathbf{A}^R$ and
entity attribute matrix {\small$\mathbf{X}$}:

\begin{small}
\begin{flalign}\label{eq:adj_matrix3}
\mathbf{A}^R\mathbf{X}=\left[\begin{array}{cc}
\mathbf{0}^{M\times D_u} & \mathbf{R}\mathbf{X}^V\\
\mathbf{R}^T\mathbf{X}^U & \mathbf{0}^{N\times D_v}
\end{array}\right]
=\left[\begin{array}{cc}
\mathbf{0}^{M\times D_u} & \mathbf{S}_v\\
\mathbf{S}_u & \mathbf{0}^{N\times D_v}
\end{array}\right].
\end{flalign}
\end{small}

where {\small$\mathbf{X}^U$} represents the user attribute part of {\small$\mathbf{X}$} and {\small$\mathbf{X}^V$} represents the item attribute part of {\small$\mathbf{X}$}.
Since {\small$\mathbf{S}_u$} is a two-order link matrix, in which $s_{jk} \geq 1$ indicates the count that item $j$ has indirect links with user attribute $k$. $s_{jk} = 0$ denotes there is no indirect link between item $j$ and user attribute $k$.
The user student model's graph structure could be formulated as a item-user attribute graph: $\mathcal{G}_{S_u} = <V \cup \mathbf{X}^U, \mathbf{A}^{S_u}>$,  where the graph adjacent matrix is constructed from the item-user attribute matrix {\small$\mathbf{A}^{S_u}$}:

\begin{small}
\begin{flalign}\label{eq:adj_matrix4}
\mathbf{A}^{S_u}=\left[\begin{array}{cc}
\mathbf{0}^{N\times N} & \mathbf{S}_u\\
\mathbf{S}_u^T & \mathbf{0}^{D_u\times D_u}
\end{array}\right].
\end{flalign}
\end{small}

Since this student graph $\mathcal{G}_{S_u}$ is constructed based on second-order connections, it will be a little denser than traditional user-item graph.
After graph construction, we employ the item embedding from the teacher model as the initial embedding of the item in the student model. 
For the user attribute embedding $\mathbf{e}_k \in \mathbb{R}^{d}$, since user attributes only have indirect connection with items, we do not employ the user attribute embedding from teacher model and initialize it with Gaussian Distribution on the other hand. 

\textbf{Embedding Propagation Layer.}
Since there only exist indirect links between items and user attributes, we leverage the item free embedding to update the attribute embedding $\mathbf{e}_k$. 
Taking the update in the $(t+1)^{th}$ layer as an example, we aggregate the item neighbors of user attribute $k$ to update its embedding.
Let ${A_{k+N}^{S_u}} = \{j|s_{jk} \geq 1 \}$ denotes the item set that has indirect connection with user attribute $k$, the $(t+1)^{th}$ updating operation can be formulated as follows:

\begin{flalign}\label{eq:embedding propagation3}
\mathbf{e}_k^{t+1} &= (\mathbf{e}_k^{t}+\sum_{j\in {A_{k+N}^{S_u}}} \frac{\mathbf{v}_j^{t}}{|A_{k+N}^{S_u}|}).
\end{flalign}

Meanwhile, item embedding can be updated with the corresponding user attribute neighbors in a similar way.
Let ${A_j^{S_u}} = \{k|s_{jk} >=1 \}$ denotes the user attribute set that has indirect connections with item $j$. The $(t+1)^{th}$ updating operation can be described as follows:

\begin{flalign}\label{eq:embedding propagation4}
\mathbf{v}_j^{t+1} &= (\mathbf{v}_j^{t}+\sum_{j\in {A_j^{S_u}}} \frac{\mathbf{e}_k^{t}}{|A_j^{S_u}|}).
\end{flalign}

Similar to the teacher model, let matrix  {\small$\mathbf{E}^t$}, {\small$\mathbf{V}^t$} denote the embedding matrices of user attribute in the user student model and 
items after $t^{th}$ propagation, then the updated embedding matrices after $(t+1)^{th}$ propagation as:

\begin{small}
\begin{flalign}\label{eq:new GCN1}
\left[\begin{array}{c}
\mathbf{V}^{t+1} \\
\mathbf{E}^{t+1}\end{array}\right]
=(\left[\begin{array}{c}
\mathbf{V}^{t} \\
\mathbf{E}^{t}\end{array}\right]+
{\mathbf{D}^{S_u}}^{-1}\mathbf{A}^{S_u}\times
\left[\begin{array}{c}
\mathbf{V}^{t} \\
\mathbf{E}^{t}\end{array}\right]).
\end{flalign}
\end{small}

Finally, we can get the user attribute embedding and the updated item free embedding. Taking new user $i$ and item $j$ as an example, the attribute set of new user $i$ can be represented with $X_{u_i} = \{k|\bar{x}_{ik} = 1\}$. Their embeddings can be represented as follows:
\begin{equation}
    \begin{split}
    \mathbf{u}_i^U = \sum_{k \in X_{u_i}}\mathbf{e}_k^{L_{S_u}}, \quad 
    \mathbf{v}_j^U = \mathbf{v}_j^{L_{S_u}},
    \end{split}
\end{equation}
where $L_{S_u}$ is the number of GCN layers in the user student model. 
Meanwhile, we can obtain the user embedding $\mathbf{u}_i^I$ and item embedding $\mathbf{v}_j^I$ in a similar way.

\textbf{Prediction Layer.}
In this layer, we intend to utilize the learned user embedding and item embedding to calculate the corresponding rating. Taking user $i$ and item $j$ as an example, the predicted rating can be calculated with the following function:
\begin{equation}
\label{eq:overall_prediction}
    \begin{split}
        \hat{r}_{ij} = \hat{\mathbf{u}}_i(\hat{\mathbf{v}}_j)^T.
    \end{split}
\end{equation}

If the user and item are available simultaneously, the predicted rating can be obtained with $\hat{\mathbf{u}}_i=\mathbf{u}_i^L, \hat{\mathbf{v}}_j=\mathbf{v}_j^L$, as illustrated in Eq.\ref{eq:teacher_predict}. 
When dealing with cold start problem,
we employ different components in \model~to generate different implementations of user embedding $\hat{\mathbf{u}}_i$ and item embedding $\hat{\mathbf{v}}_j$ in Eq.~\ref{eq:overall_prediction}, which is in favor of tackling different situations of cold start problem in a unified way.

\textit{1) Task 1.}
In this task, we select the user student model.
User embedding $\mathbf{u}_i$ can be represented with the sum of corresponding attribute embedding $\mathbf{e}_k (k\in{X_{u_i}})$ in user student model. 
Item embedding can be represented with the free embedding $\mathbf{v}_j^L$ generated in teacher
model. Finally, Eq.~\ref{eq:overall_prediction} can be modified as follows:
\begin{equation}
\begin{split}
    \hat r_{ij} = \hat{\mathbf{u}}_i(\hat{\mathbf{v}}_j)^T = \mathbf{u}_i^U(\mathbf{v}_j^L)^T
                = (\sum_{k \in X_{u_i}}\mathbf{e}_k^{L_{S_u}})(\mathbf{v}_j^L)^T.
\end{split}
\end{equation}

\textit{2) Task 2.}
In this task, we select the item student model. 
For user embedding, we select the user free embedding $\mathbf{u}_i^L$ from Teacher model as the representation.
For item representation, we make full use of its attribute embedding $f_l (l\in{X_{v_j}})$ as the needed embedding. 
Therefore, Eq.~\ref{eq:overall_prediction} is modified as follows:
\begin{equation}
\begin{split}
    \hat r_{ij} = \hat{\mathbf{u}}_i(\hat{\mathbf{v}}_j)^T = \mathbf{u}_i^L(\mathbf{v}_j^I)^T
                = (\mathbf{u}_i^L)(\sum_{l \in X_{v_j}}\mathbf{f}_l^{L_{S_v}})^T.
\end{split}
\end{equation}

\textit{3) Task 3.}
In this task, the user and item free embedding are not available at the same time. 
Therefore, we employ both user student model and item student model to generate the user and item embeddings with their attribute information. 
Specifically, we select the user embedding $\mathbf{u}_i^U$ and item embedding $\mathbf{v}_j^I$, which are driven from their own attributes, and modify Eq.~\ref{eq:overall_prediction} as follows:
\begin{equation}
\begin{split}
    \hat r_{ij} = \hat{\mathbf{u}}_i(\hat{\mathbf{v}}_j)^T = \mathbf{u}_i^U(\mathbf{v}_j^I)^T
                = (\sum_{k \in X_{u_i}}\mathbf{e}_k^{L_{S_u}})(\sum_{l \in X_{v_j}}\mathbf{f}_l^{L_{S_v}})^T.
\end{split}
\end{equation}

\subsection{Model Optimization}
Since \model~contains two main components, the optimization also consists of two parts: 
\textit{Rating Prediction Loss} for Teacher Model, and \textit{Graph Distillation Loss} for \model.

\textbf{Rating Prediction Loss.}
For recommender system based on implicit feedback, BPR-based on pair-wise ranking is the most popular optimization algorithm. 
Thus, the objective function can be formulated as follows:
\begin{align}
    L_r = \sum_{u\in U}\sum_{(i,j)\in B_u}-ln\sigma(\hat{r}_{ui} - \hat{r}_{uj}) + \gamma ||\theta||^2,
\end{align}
where $\sigma(\cdot)$ is a sigmoid activation function. 
$B_u=\{(i,j)|r_{ui}=1\!\wedge\!r_{uj}\neq 1\}$ denotes the pairwise training data for user $u$. $\hat{r}_{ui}$ and $\hat{r}_{uj}$ are computed by the free embedding of
the teacher model.
$\theta$ represents the user and item free embedding matrices. 
$\gamma$ is a regularization parameter that restrains the user and item free latent embedding matrices.

\textbf{Graph Distillation Loss.} 
Since distillation techniques are employed in \model~to help the student model to learn better user and item embeddings, as well as make accurate predictions based on the attribute information,  with the guidance of teacher model. 
Thus, the learned user embedding $\mathbf{u}_i^L$~(item embedding $\mathbf{v}_j^L$) from teacher model and $\mathbf{u}_i^U$~($\mathbf{v}_j^I$) from student model should be similar.
This optimizing target can be formulated as follows:
\begin{equation}
    \begin{split}
        L_u = \sum_{i=0}^{M-1} ||\mathbf{u}_i^L-\mathbf{u}_i^U||^2, \quad
        L_v = \sum_{j=0}^{N-1} ||\mathbf{v}_j^L-\mathbf{v}_j^I||^2.
    \end{split}
\end{equation}

Meanwhile, we intend the student model to predict user preference correctly. 
{\small$\mathbf{U}$} and {\small$\mathbf{V}$} represent the embedding matrices of users and items in the teacher model.
{\small$\mathbf{U}^U$} and {\small$\mathbf{V}^I$} represent the embedding matrices of users and items in the student model.
Thus, its prediction result should be similar to the results of the teacher model.
which can be formulated as follows:

\begin{small}
\begin{equation}
    \begin{split}
        L_s = ||\mathbf{U}\mathbf{V}^T-\mathbf{U}^U(\mathbf{V}^I)^T||^2.
    \end{split}
\end{equation}
\end{small}
The Graph Distillation Loss will be formulated as follows:
\begin{small}
\begin{equation}
\label{eq:distillation_eq}
    \begin{split}
        L_d = \lambda L_u + \mu L_v + \eta L_s,
    \end{split}
\end{equation}
\end{small}
where $\lambda, \mu, \eta$ are the weight of different information distillation loss. 
We can adjust their values to focus our proposed \model~on tackling different sub-tasks in code start problem in recommendation. 
After obtaining the two parts objective functions, The final optimization of our model can be formulated as follows:
\begin{align}
    Loss = L_{r} + L_{d}
\end{align}

\section{Experiments}
\label{s:experiment}
In this section, we conduct extensive experiments on three datasets to verify the effectiveness of our proposed \model~for cold start recommendation. 
We aim to answer the following questions:
\begin{itemize}
    \item Will the attribute information and the utilization method in \model~be useful for solving the cold start problem~(e.g., new users or new items) in recommendations?
    
    \item Is the distillation technique helpful for student model to learn useful knowledge from teacher model for user or item embedding? 
    
    \item What is the influence of each component in our proposed \model~to the overall performance?
\end{itemize}

\subsection{Datasets}

In this paper, we select three suitable and public available datasets to evaluate all the models, i.e., Yelp, XING~\cite{abel2017recsys}, and Amazon-Video Games~\cite{he2016vbpr}. 
Table~\ref{tab:statistics} report the statistics of three datasets.

\begin{table}[]
\caption{The statistics of the three datasets.}\label{tab:statistics}
\vspace{-0.2cm}
\scalebox{0.85}{
\begin{tabular}{|l|l|c|c|c|}
\hline
\multicolumn{2}{|c|}{Dataset}                                                                     & Yelp    & XING    & \begin{tabular}[c]{@{}c@{}}Amazon-\\ Video Games\end{tabular} \\ \hline
\multicolumn{1}{|c|}{\multirow{4}{*}{Train}}                                          & Old Users & 29,777  & 20,640  & 29,129                                                        \\ \cline{2-5} 
\multicolumn{1}{|c|}{}                                                                & Old Items & 27,737  & 17,793  & 22,547                                                        \\ \cline{2-5} 
\multicolumn{1}{|c|}{}                                                                & Ratings   & 159,857 & 133,139 & 172,089                                                       \\ \cline{2-5} 
\multicolumn{1}{|c|}{}                                                                & Density   & 0.019\% & 0.036\% & 0.026\%                                                       \\ \hline
\multicolumn{1}{|c|}{\multirow{3}{*}{Val}}                                            & Old Users & 2,109   & 17,058  & 26,506                                                        \\ \cline{2-5} 
\multicolumn{1}{|c|}{}                                                                & Old Items & 1,812    & 10,357  & 10,189                                                        \\ \cline{2-5} 
\multicolumn{1}{|c|}{}                                                                & Ratings   & 2,109   & 20,258  & 29,870                                                        \\ \hline
\multirow{3}{*}{Test new user}                                                        & New Users & 12,749  & 7,105   & /                                                             \\ \cline{2-5} 
                                                                                      & Old Items & 17,121  & 7,665   & /                                                             \\ \cline{2-5} 
                                                                                      & Ratings   & 65,127  & 12,858  & /                                                             \\ \hline
\multirow{3}{*}{Test new item}                                                        & Old Users & 27,067  & 11,013  & 22,027                                                        \\ \cline{2-5} 
                                                                                      & New Items & 11,975  & 7,598   & 10,170                                                        \\ \cline{2-5} 
                                                                                      & Ratings   & 69,524  & 33,079  & 98,044                                                        \\ \hline
\multirow{3}{*}{\begin{tabular}[c]{@{}l@{}}Test new user\\ and new item\end{tabular}} & New Users & 11,662  & 4,618   & /                                                             \\ \cline{2-5} 
                                                                                      & New Items & 8,734   & 4,276   & /                                                             \\ \cline{2-5} 
                                                                                      & Ratings   & 30,288  & 7,318    & /                                                             \\ \hline
\multicolumn{2}{|l|}{User Attributes}                                                    & 80      & 108     & /                                                             \\ \hline
\multicolumn{2}{|l|}{Item Attributes}                                                    & 183     & 81      & 76                                                            \\ \hline
\end{tabular}
}
\vspace{-4mm}
\end{table}

In order to evaluate the model performance on each of three sub-tasks in cold start problem, we manually set the new users or new items in the test sets\cite{zhu2020recommendation}.
Specifically, we randomly select $30\%$ users in the test set.
Then, we keep the corresponding items and remove their connections to construct the new user test set for Task 1. 
Meanwhile, we apply the same operation to generate a new items test set for Task 2.
As for Task 3, we collated interaction records belonging to both the new user and the new product as the test set. Then, we split 10\% validation set from the  rest old users and old items.
The details are reported in Table~\ref{tab:statistics}.

\subsection{Experimental Setup}

\textbf{Evaluation Metrics.}
Since the cold start problem still can be treated as top-K recommendation task, we select two popular ranking metrics to evaluate our model: HR@K and NDCG@K ($K=\{10,20,50\}$).

\begin{table*}[]
\setlength{\belowcaptionskip}{2pt}
\textbf
{\caption{HR@K and NDCG@K comparisons for Yelp and Amazon-Video Games. '-' represents unavailable result. }
\label{t:yelp_amazon_result}
}
\scalebox{0.85}{
\begin{tabular}{|c|c|c|c|c|c|c|c|c|c|c|c|c|c|}
\hline
\multirow{2}{*}{Model}        & \multicolumn{1}{l|}{\multirow{2}{*}{Metrics}} & \multicolumn{3}{c|}{Yelp(Task1)}                       & \multicolumn{3}{c|}{Yelp(Task2)}                       & \multicolumn{3}{c|}{Yelp(Task3)}                        & \multicolumn{3}{c|}{\begin{tabular}[c]{@{}c@{}}Amazon-Video Games\\ (Task2)\end{tabular}} \\ \cline{3-14} 
                              & \multicolumn{1}{l|}{}                         & @10              & @20              & @50              & @10              & @20              & @50              & @10               & @20              & @50              & @10                           & @20                         & @50                         \\ \hline
\multirow{2}{*}{KNN}          & HR                                            & 0.01810          & 0.03104          & 0.05655          & 0.01590          & 0.02775          & 0.06126          & -                 & -                & -                & 0.001270                      & 0.001898                    & 0.008407                    \\ \cline{2-14} 
                              & NDCG                                          & 0.01528          & 0.02067          & 0.02917          & 0.009864         & 0.01370          & 0.02219          & -                 & -                & -                & 0.0007551                     & 0.0009667                    & 0.002447                   \\ \hline
\multirow{2}{*}{LinMap}       & HR                                            & 0.02030          & 0.03220          & 0.05784          & 0.02011          & 0.03436          & 0.06640          & 0.01286           & 0.02480          & 0.05108          & 0.01833                       & 0.02491                     & 0.03911                     \\ \cline{2-14} 
                              & NDCG                                          & 0.01724          & 0.02231          & 0.03076          & 0.01277          & 0.01743          & 0.02561          & 0.007353          & 0.01124          & 0.01792          & 0.008335                      & 0.009481                    & 0.01333                     \\ \hline
\multirow{2}{*}{xDeepFM}      & HR                                            & 0.01984          & 0.03234          & 0.05973          & 0.02024          & 0.03491          & 0.06678          & 0.01310           & 0.02438          & 0.04898          & 0.01847                       & 0.02498                     & 0.03900                     \\ \cline{2-14} 
                              & NDCG                                          & 0.01613          & 0.02147          & 0.03054          & 0.01280          & 0.01752          & 0.02564          & 0.007516          & 0.01120          & 0.01772          & 0.008253                      & 0.009465                    & 0.01295                     \\ \hline
\multirow{2}{*}{CDL}          & HR                                            & 0.01930          & 0.03257          & 0.06041          & 0.01959          & 0.03410          & 0.06536          & 0.01268           & 0.02001          & 0.04211          & 0.02023                       & \underline{0.02775}            & \underline{0.04192}            \\ \cline{2-14} 
                              & NDCG                                          & 0.01603          & 0.02161          & 0.03082          & 0.01209          & 0.01673          & 0.02472          & \underline{0.008057} & 0.01049          & 0.01613          & 0.009470                      & \underline{0.01112}            & \underline{0.01439}            \\ \hline
\multirow{2}{*}{DropoutNet}   & HR                                            & 0.02006          & 0.03278          & 0.06029          & 0.01731          & 0.02821          & 0.05594          & 0.01143           & 0.02141          & 0.04297          & 0.01143                       & 0.01612                     & 0.02876                     \\ \cline{2-14} 
                              & NDCG                                          & 0.01675          & 0.02208          & 0.03121          & 0.01052          & 0.01411          & 0.02049          & 0.006913          & 0.009972         & 0.01547          & 0.005350                      & 0.006693                    & 0.009857                    \\ \hline
\multirow{2}{*}{Heater}       & HR                                            & \underline{0.02055} & \underline{0.03365} & 0.05880          & \underline{0.02443} & \underline{0.04179} & \underline{0.07974} & 0.01226           & 0.02440          & 0.04915          & 0.02032                       & 0.02659                     & 0.04101                     \\ \cline{2-14} 
                              & NDCG                                          & \underline{0.01726} & \underline{0.02271} & 0.03110          & \underline{0.01495} & \underline{0.02059} & \underline{0.03027} & 0.007329          & 0.01131          & \underline{0.01780} & 0.009280                      & 0.01031                     & 0.01334                     \\ \hline
\multirow{2}{*}{PinSage}      & HR                                            & 0.01985          & 0.03302          & \underline{0.06250} & 0.02080          & 0.03704          & 0.07089          & 0.01173           & 0.02110          & 0.04097          & 0.02030                       & 0.02491                     & 0.03498                     \\ \cline{2-14} 
                              & NDCG                                          & 0.01709          & 0.02267          & \underline{0.03254} & 0.01331          & 0.01856          & 0.02722          & 0.007142          & 0.01013          & 0.01523          & 0.008590                      & 0.009135                    & 0.01157                     \\ \hline
\multirow{2}{*}{PFD}          & HR                                            & 0.02015          & 0.03318          & 0.05837          & 0.02240          & 0.04008          & 0.07955          & 0.01152           & 0.02427          & 0.04766          & \underline{0.02187}              & 0.02745                     & 0.04065                     \\ \cline{2-14} 
                              & NDCG                                          & 0.01716          & 0.02247          & 0.03086          & 0.01376          & 0.01948          & 0.02952          & 0.007248          & \underline{0.01143} & 0.01758          & \underline{0.009953}             & 0.01086                     & 0.01388                     \\ \hline
\multirow{2}{*}{Student}      & HR                                            & 0.01886          & 0.03133          & 0.05944          & 0.02290          & 0.03984          & 0.07625          & \underline{0.01317}  & \underline{0.02540} & \underline{0.05109} & 0.01812                       & 0.02328                     & 0.03457                     \\ \cline{2-14} 
                              & NDCG                                          & 0.01612          & 0.02140           & 0.03074          & 0.01419          & 0.01968          & 0.02897          & 0.007294          & 0.01122          & 0.017952         & 0.008275                      & 0.009040                     & 0.01205                     \\ \hline
\multirow{2}{*}{\textbf{\model}} & \textbf{HR↑}                                  & \textbf{0.02077} & \textbf{0.03404} & \textbf{0.06426} & \textbf{0.02717} & \textbf{0.04712} & \textbf{0.08856} & \textbf{0.01443}  & \textbf{0.02589} & \textbf{0.05117} & \textbf{0.02240}              & \textbf{0.02953}            & \textbf{0.04507}            \\ \cline{2-14} 
                              & \textbf{NDCG↑}                                & \textbf{0.01767} & \textbf{0.02323} & \textbf{0.03324} & \textbf{0.01659} & \textbf{0.02306} & \textbf{0.03366} & \textbf{0.008653} & \textbf{0.01240} & \textbf{0.01890} & \textbf{0.01008}              & \textbf{0.01164}            & \textbf{0.01601}            \\ \hline
\end{tabular}}
\end{table*}

\begin{table*}[]
\setlength{\belowcaptionskip}{2pt}
    \textbf
    {\caption{HR@K and NDCG@K comparisons for XING. '-' represents unavailable result. KNN cannot work for task3.}
    \label{t:xing_result}
    }
\scalebox{0.95}{
\begin{tabular}{|c|c|c|c|c|c|c|c|c|c|c|}
\hline
\multirow{2}{*}{Model}        & \multicolumn{1}{l|}{\multirow{2}{*}{Metrics}} & \multicolumn{3}{c|}{XING(Task1)}                         & \multicolumn{3}{c|}{XING(Task2)}                          & \multicolumn{3}{c|}{XING(Task3)}                          \\ \cline{3-11} 
                              & \multicolumn{1}{l|}{}                         & @10               & @20               & @50              & @10               & @20               & @50               & @10               & @20               & @50               \\ \hline
\multirow{2}{*}{KNN}          & HR                                            & 0.002977          & 0.005945          & 0.01249          & 0.001345          & 0.002246          & 0.005768          & -                 & -                 & -                 \\ \cline{2-11} 
                              & NDCG                                          & 0.001586          & 0.002436          & 0.003920         & 0.0006946         & 0.0009711         & 0.001913          & -                 & -                 & -                 \\ \hline
\multirow{2}{*}{LinMap}       & HR                                            & 0.007926          & 0.01483           & 0.02628          & 0.002039          & 0.003692          & 0.007225          & 0.001552          & 0.003338          & 0.007983          \\ \cline{2-11} 
                              & NDCG                                          & 0.004242          & 0.006225          & 0.008781         & 0.001047          & 0.001559          & 0.002492          & 0.0007650         & 0.001291          & 0.002255          \\ \hline
\multirow{2}{*}{xDeepFM}      & HR                                            & 0.007733          & 0.01530           & 0.02752          & 0.001991          & \underline{0.003892}    & 0.007474          & 0.002526          & 0.005242          & \underline{0.009932}    \\ \cline{2-11} 
                              & NDCG                                          & 0.004240          & 0.006289          & 0.009048         & 0.0009840         & 0.001526          & 0.002450          & 0.0009600         & 0.001765          & \underline{0.002794}    \\ \hline
\multirow{2}{*}{CDL}          & HR                                            & 0.007546          & 0.01469           & 0.02815          & 0.001521          & 0.003213          & 0.006708          & \underline{0.002992}    & 0.004580           & 0.007668          \\ \cline{2-11} 
                              & NDCG                                          & 0.004250          & 0.006255          & 0.009263         & 0.0008030         & 0.001357          & 0.002334          & \underline{0.001479}    & 0.001854          & 0.002444          \\ \hline
\multirow{2}{*}{DropoutNet}   & HR                                            & 0.006997          & 0.01278           & 0.02345          & 0.001404          & 0.003784          & 0.007138          & 0.001805          & 0.003901          & 0.007610          \\ \cline{2-11} 
                              & NDCG                                          & 0.003311          & 0.004959          & 0.007376         & 0.0007770         & 0.001531          & 0.002458          & 0.0008680         & 0.001332          & 0.002222          \\ \hline
\multirow{2}{*}{Heater}       & HR                                            & 0.006934          & 0.01524           & 0.02717          & 0.001766          & 0.003633          & 0.007661          & 0.002635          & 0.004788          & 0.007963          \\ \cline{2-11} 
                              & NDCG                                          & 0.003354          & 0.005713          & 0.008451         & 0.001061          & \underline{0.001667}    & \underline{0.002722}    & 0.001429          & 0.001704          & 0.002415          \\ \hline
\multirow{2}{*}{PinSage}      & HR                                            & 0.004862          & 0.01119           & 0.02193          & 0.001646          & 0.003693          & \underline{0.007953}    & 0.001002          & 0.002315          & 0.003741          \\ \cline{2-11} 
                              & NDCG                                          & 0.002680          & 0.004436          & 0.006818         & 0.0009460         & 0.001610          & 0.002705          & 0.0004690         & 0.001046          & 0.001358          \\ \hline
\multirow{2}{*}{PFD}          & HR                                            & \underline{0.009043}    & 0.01552           & 0.02855          & \underline{0.002331}    & 0.003877          & 0.007373          & 0.002833          & \underline{0.005251}    & 0.008742          \\ \cline{2-11} 
                              & NDCG                                          & \underline{0.005273}    & \underline{0.007073}    & 0.01005          & \underline{0.001151}    & 0.001666          & 0.002578          & 0.001300          & \underline{0.001942}    & 0.002695          \\ \hline
\multirow{2}{*}{Student}      & HR                                            & 0.008985          & \underline{0.01725}     & \underline{0.03114}    & 0.001998          & 0.003734          & 0.007789          & 0.001777          & 0.004040           & 0.006881          \\ \cline{2-11} 
                              & NDCG                                          & 0.004734          & 0.007033          & \underline{0.01015}    & 0.0009520         & 0.001460          & 0.002506          & 0.0008830          & 0.001526          & 0.002144          \\ \hline
\multirow{2}{*}{\textbf{\model}} & \textbf{HR↑}                                  & \textbf{0.01149}  & \textbf{0.02204}  & \textbf{0.04060} & \textbf{0.002539} & \textbf{0.004216} & \textbf{0.008276} & \textbf{0.003999} & \textbf{0.006727} & \textbf{0.01018}  \\ \cline{2-11} 
                              & \textbf{NDCG↑}                                & \textbf{0.006522} & \textbf{0.009160} & \textbf{0.01330} & \textbf{0.001322} & \textbf{0.001758} & \textbf{0.002780} & \textbf{0.001694} & \textbf{0.002222} & \textbf{0.002886} \\ \hline
\end{tabular}}
\end{table*}

\begin{table*}[]
\setlength{\belowcaptionskip}{2pt}

\caption{HR@20 and NDCG@20 results of our model with different propagation depth $L$ on Yelp and Amaon-Video Games (We fix the same gcn layer $L$ of student model and teacher model).}
\label{t:yelp_amazon_gcn_layer}
\scalebox{0.90}{
\begin{tabular}{|c|r|r|r|r|r|r|r|r|}
\hline

\multirow{2}{*}{Num. of GCN Layers} & \multicolumn{2}{c|}{Yelp(Task1)}                          & \multicolumn{2}{c|}{Yelp(Task2)}                          & \multicolumn{2}{c|}{Yelp(Task3)}                          & \multicolumn{2}{c|}{Amazon Video Games}                   \\ \cline{2-9} 
                                    & \multicolumn{1}{c|}{HR@20} & \multicolumn{1}{c|}{NDCG@20} & \multicolumn{1}{c|}{HR@20} & \multicolumn{1}{c|}{NDCG@20} & \multicolumn{1}{c|}{HR@20} & \multicolumn{1}{c|}{NDCG@20} & \multicolumn{1}{c|}{HR@20} & \multicolumn{1}{c|}{NDCG@20} \\ \hline
$L=1$                                   & 0.03365                   & 0.02294                      & 0.04541                    & 0.02239                     & 0.02467                   & 0.01142                   & 0.02946                    & 0.01125                      \\ \hline
$L=2$                                   & \textbf{0.03404}           & \textbf{0.02323}             & 0.04606                  & 0.02255                   & \textbf{0.02589}           & \textbf{0.01240}             & \textbf{0.02953}           & \textbf{0.01164}             \\ \hline
$L=3$                                 & 0.03355                   & 0.02186                    & \textbf{0.04712}           & \textbf{0.02306}             & 0.02577                    & 0.01198                     & 0.02801                    & 0.01124                      \\ \hline
$L=4$                                 & 0.03225                    & 0.02102                      & 0.04693                   & 0.02298                      & 0.02533                    & 0.01192                      & 0.02707                    & 0.01104                      \\ \hline
\end{tabular}
}
\end{table*}

\begin{table*}[]
\setlength{\belowcaptionskip}{2pt}
\caption{HR@20 and NDCG@20 results of our model with different propagation depth $L$ on XING (We fix the same gcn layer $L$ of student model and teacher model).
\label{t:xing_gcn_layer}} 
\scalebox{1}{
\begin{tabular}{|c|r|r|r|r|r|r|}                             \hline
                                     & \multicolumn{2}{c|}{XING(Task1)}                          & \multicolumn{2}{c|}{XING(Task2)}                          & \multicolumn{2}{c|}{XING(Task3)}                                      \\ \cline{2-7} 
\multirow{-2}{*}{Num. of GCN Layers} & \multicolumn{1}{c|}{HR@20} & \multicolumn{1}{c|}{NDCG@20} & \multicolumn{1}{c|}{HR@20} & \multicolumn{1}{c|}{NDCG@20} & \multicolumn{1}{c|}{HR@20}             & \multicolumn{1}{c|}{NDCG@20} \\ \hline
$L=1$                                    & 0.02071                  & 0.008274                     & 0.004003                   & 0.001754                     & 0.006107                               & 0.001962                    \\ \hline
$L=2$                                    & 0.02107                    & 0.009037                   & \textbf{0.004216}          & \textbf{0.001758}            & 
             \textbf{0.006727} & \textbf{0.002222}             \\ \hline
$L=3$                                    & \textbf{0.02204}           & \textbf{0.009160}            & 0.003992                   & 0.001752                     & 0.006439                               & 0.002100                     \\ \hline
$L=4$                                    & 0.02176                  & 0.008947                    & 0.003907                   & 0.001672                     & 0.006359                             & 0.002054                    \\ \hline
\end{tabular}}
\end{table*}

\begin{figure*}
\vspace{-0.2cm}
  \vspace{-0.2cm}
      \subfigure[Varying  \(\lambda\) in Yelp (Task 1)]{\includegraphics[width=52mm]{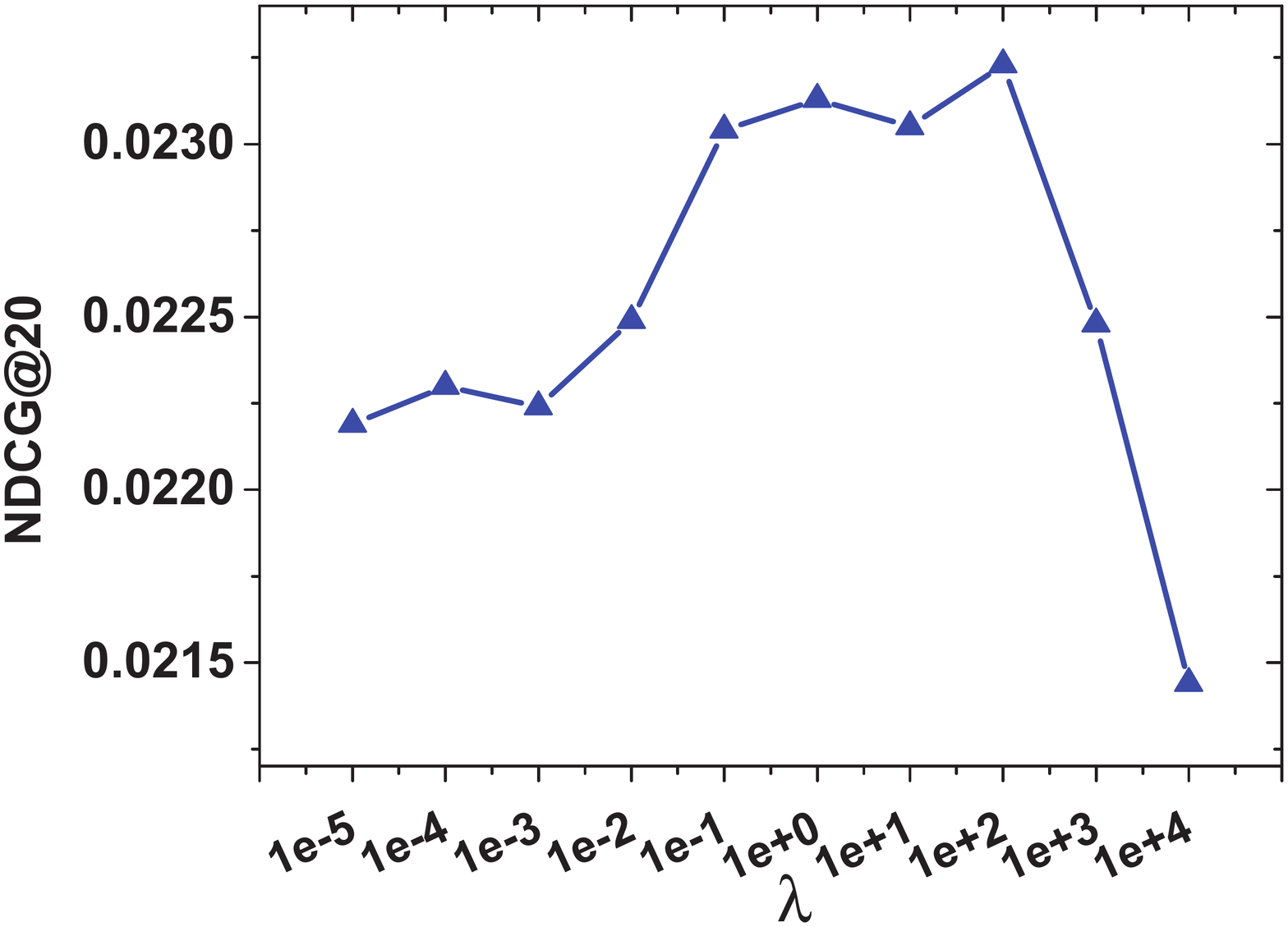}}
      \subfigure[Varying  \(\mu\) in Yelp (Task 2)]{\includegraphics[width=52mm]{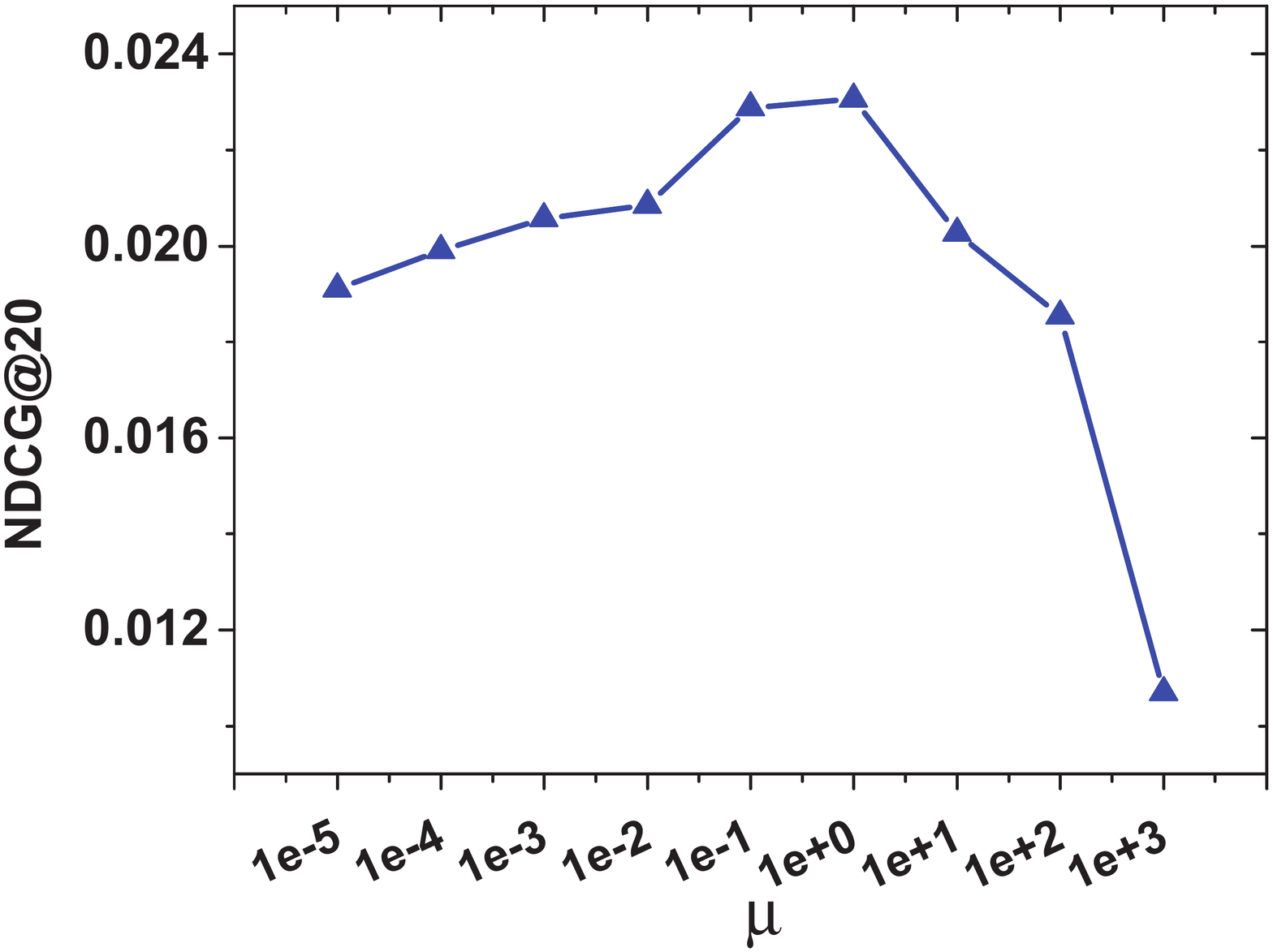}}
      \subfigure[Varying  \(\eta\) in Yelp (Task 3)]{\includegraphics[width=52mm]{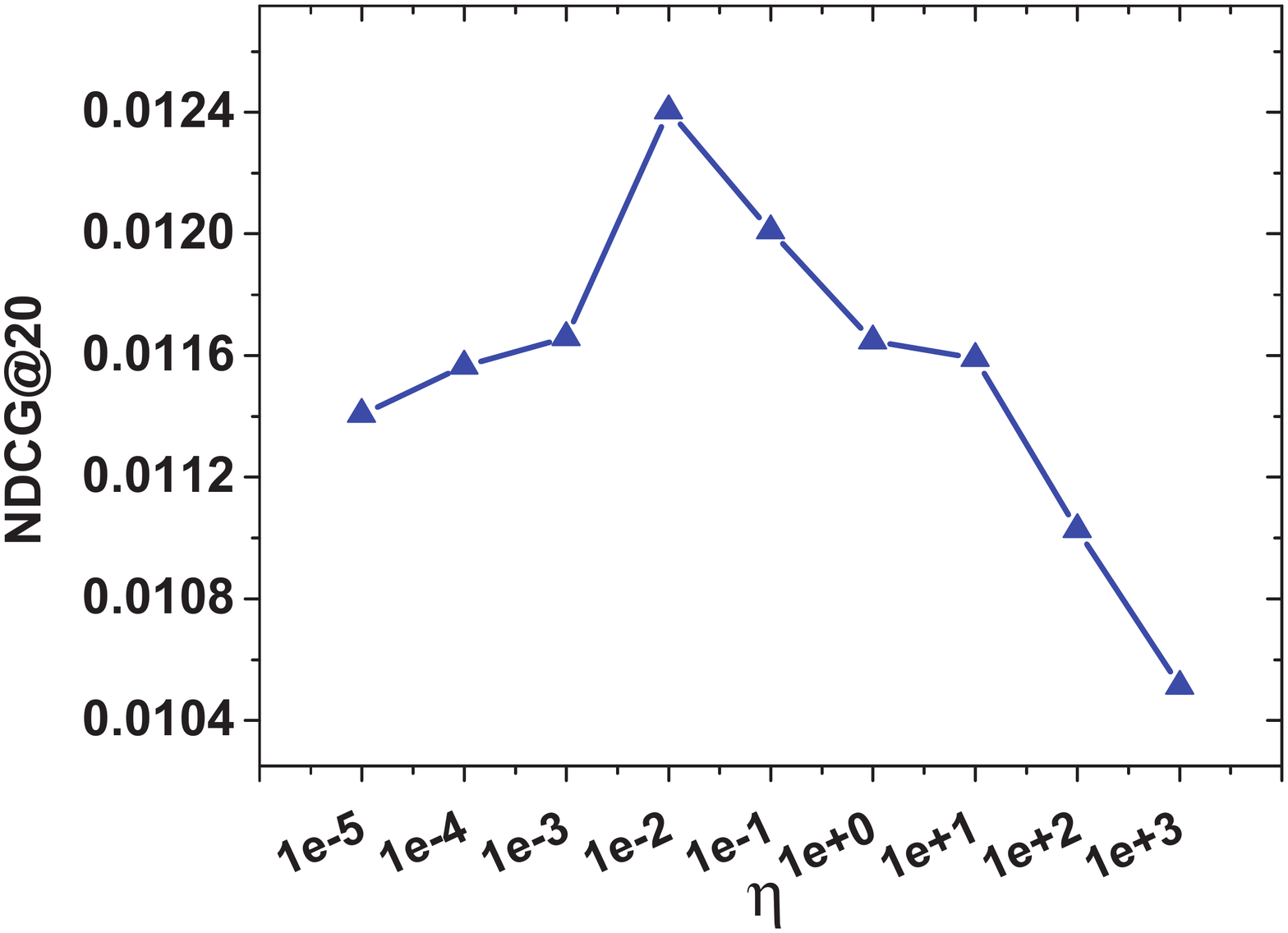}}
  \vspace{-0.2cm}
  \caption{NDCG@20 results of our model with different hyper-parameters.} \label{fig:hyper_parameters_results}
  \vspace{-0.2cm}
\end{figure*}

\begin{table}[]
\setlength{\belowcaptionskip}{2pt}
\caption{HR@20 and NDCG@20 results of output distillation and multi-layer distillation.
\label{t:output_multi_layer}} 
\scalebox{0.95}{
\begin{tabular}{|c|c|c|c|}
\hline
\multicolumn{1}{|l|}{}       & Metrics & 2-Layer Output  & 2-Layer Multi-Layer \\ \hline
\multirow{2}{*}{Yelp(Task1)} & HR@20   & 0.03404         & \textbf{0.03415}     \\ \cline{2-4} 
                             & NDCG@20 & 0.02323         & 0.02152              \\ \hline
\multirow{2}{*}{Yelp(Task2)} & HR@20   & 0.04606         & \textbf{0.04627}     \\ \cline{2-4} 
                             & NDCG@20 & 0.02255         & \textbf{0.02302}     \\ \hline
\multirow{2}{*}{Yelp(Task3)} & HR@20   & 0.02589         & \textbf{0.02638}     \\ \cline{2-4} 
                             & NDCG@20 & 0.01240         & 0.01205              \\ \hline
\multirow{2}{*}{Amazon}      & HR@20   & 0.02953         & 0.02812              \\ \cline{2-4} 
                             & NDCG@20 & 0.01164         & 0.01113              \\ \hline
\end{tabular}
}
\end{table}

\textbf{Parameter Settings.}
First of all, the dimensions of collaborative filtering embedding and the attribute representation are all set as $64$.
The batch size is set as $2,048$. 
The depth $L$ of GCN is selected from $\{1,2,3,4\}$, and we also make an experiment to verify the influence of different depths.
During training, Adam is employed as the optimizer with learning rate $0.001$. 

Gaussian distribution with a mean of 0 and variance of 0.01 is employed to initialize the embedding matrices. At each iteration of the training process, we randomly sample one candidate negative sample to
compose a triple data. In the testing phase, to avoid the
unfairness caused by the randomly negative samples, we evaluted
all models in the condition of all negative samples.
As shown in Eq.~\ref{eq:distillation_eq}, there are three hyper-parameters $\lambda, \mu$ and $\eta$. 
We tune the three hyper-parameters on three different tasks respectively.
The combination for Yelp is $\{\lambda=100, \mu=1, \eta=0.01\}$, for Amazon-Video Games is $\{\mu=10\}$ and for XING is $\{\lambda=1, \mu=100, \eta=0.001\}$.

\subsection{Overall Results}
Tables~\ref{t:yelp_amazon_result} and \ref{t:xing_result} report the overall results on three datasets. 
We can obtain that  \model~outperforms all baselines across all the datasets with different evaluation metrics. Specifically, \model~achieves average {$2.03\%$, $11.67\%$, $6.01\%$} improvement across three sub-tasks on Yelp, average {$27.83\%$, $7.28\%$, $16.06\%$} improvement on XING, and average {$5.6\%$} improvement on Amazon-Game Videos, respectively. This phenomenon demonstrates the effectiveness of introducing attribute information into graph as node and learning attribute embedding and entity embeddings simultaneously under the graph constraint. 
Moreover, \model~makes full use of distillation techniques to narrow down the gap between attribute embedding and CF-based embedding and help the student model to learn entity embedding from the teacher model with the attribute information as input. 

Meanwhile, \model~tries to tackle all three sub-tasks in a unified framework. 
To this end, we also designed a student baseline to address new item or new user problem independently.  
Specifically, for Task 1, we only select the user student model to learn the user attribute embedding and item CF-based embedding. For Task 2, we have similar operations.  As for Task 3, we select the user attribute embedding and item attribute embedding from two student models. 
The corresponding results are illustrated in Tables~\ref{t:yelp_amazon_result} and \ref{t:xing_result}. 
We can obtain that \model~still outperforms the student baselines, indicating the superiority and necessity of distilling and modeling user preference in a unified way.

\subsection{The Impact of Different Propagation Layer Depth L and Detailed Model Analysis. }
As introduced in Section~\ref{s:model}, the number of GCN layers will has a big impact on the model performance. 
Therefore, we conduct additional experiments to verify its impact. 
Corresponding results are illustrated in Tables~\ref{t:yelp_amazon_gcn_layer} and \ref{t:xing_gcn_layer}. 
From the results, we can obtain that with the increasing number of GCN layers in the teacher model, the overall performance first rises and then falls. 
When the number of GCN layers is 2 or 3, \model~achieved the best performance. 
The possible reason is that with the increasing number of GCN layers, each node could aggregate more neighbors' information, which not only alleviate the data sparsity problem, but also gather more useful information for node embedding learning. 
On the other hand, too many GCN layers in the teacher model will cause the student hard to follow and node feature over smoothing problem. 
Therefore, we select $2$ or $3$ as the GCN layer number in teacher model according to tasks and datasets.

The above analysis shows that \model~can distill knowledge at the output layer.
Intuitively, applying distillation operations to each layer seems to get better performance.
We conduct experiments to compare the effects of the two distillation methods in Table~\ref{t:output_multi_layer}. 
At 2 layer, multi-layer distillation has a little improved effect on task2 of the yelp dataset. 
However, there is no general enhancement but still competitive against baselines on the other tasks.
We speculate the reason is that, there is still a gap between the intermediate layer embedding distillation and the final output embedding distillation.
Our model is not a direct node-to-node distillation between teacher graph and student graph,
and the final entity embedding of the student model fuses the attribute node information.
Multi-layer distillation only relies on the weighted sum operation which does not capture well the positive impact of the distillation of the first layer on the final output distillation.

\subsection{Ablation Study}
In the previous parts, we have illustrated the superiority of our proposed \model. 
However, the student model tries to distill knowledge from teacher model with three constraints~(i.e., user embedding constraint, item embedding constraint, and prediction constraint), which component plays a more important role in user preference modeling is still unclear. 
To this end, we conduct an ablation study on parameters $\{\lambda,\mu,\eta\}$ to verify the impact of each component with NDCG@20. 
When verifying the effectiveness of one constraint, we fix other two parameters and modify the corresponding weight to obtain the results.
Figure~\ref{fig:hyper_parameters_results} reports the corresponding results, from which we can obtain the following observations.

With the increase of each component, model performance first increases and then decreases. 
The distillation loss constraint has a negative impact on the teacher model when the distillation loss is overweight.  
Moreover, when \model\ achieves the best performance, $\lambda$ and $\mu$ have similar values.
Thus, we can conclude that the user embedding constraint and item embedding constraint have similar impacts on model performance. 
Furthermore, we can observe that the best value for $\eta$ is very small. 
Since this is a top-K recommendation task, the prediction constraint may have a big impact on the final performance.

We also observed that the boosting effect of these parameters is different for different tasks and different datasets. For instance, the metrics of task1 in Yelp improved $2.03\%$, but improved $11.67\%$ of task2 in Yelp. 
We speculate the possible reason is that the types of user attributes are less than item attributes.
Thus, user attributes cannot provide as much information as item attributes do. 
Therefore, the user embedding distillation may not be as good as item embedding distillation. 
As a result, item embedding constraint has a bigger impact on the model performance.

\section{CONCLUSION}
In this paper, we argued that attribute information is not fully explored in cold start recommendations.
Thus, we proposed a novel \fmodel~to constrain the attribute embedding and CF-based embedding learning in a graph manner and leverage distillation technique to tackle the cold start recommendation.
In concerned details, we first introduce attributes as nodes into user-item graph and learn attribute embedding and CF-based embedding simultaneously. 
Then, we employed distillation technique to guide \model~to learn the transformation between CF-based embedding and attribute embedding. 
Thus, the student model can learn effective user (item) embedding based on attribute information from the teacher model. 
Extensive experiments on three
public datasets show the performance improvement of \model~over state-of-the-art baselines. 
In the future, we plan to explore different distillation architectures to better attribute node embedding.

\vspace{-0.1cm}

\section*{Acknowledgements}

This work was supported in part by grants from the National Natural Science Foundation of China (Grant No. U1936219, U19A2079, 62006066, 61932009), the Young Elite Scientists Sponsorship Program by CAST and ISZS, CCF-Tencent RAGR20200121, and the Open Project Program of the National Laboratory of Pattern Recognition (NLPR).

\bibliographystyle{ACM-Reference-Format}
\bibliography{bibi}

\appendix

\end{document}